\documentclass[aps,pre,preprint,superscriptaddress,floatfix]{revtex4-2}
\usepackage[utf8]{inputenc}
\usepackage{graphicx}
\usepackage{dcolumn}% Align table columns on decimal point
\usepackage{bm}% bold math
\graphicspath{{../pdf/}{D:\ImagesforProjectLatex}}
\usepackage{caption} % for \captionof
\usepackage{textcomp}
\usepackage{float}
\usepackage{epstopdf}
\usepackage{xcolor}
\usepackage{verbatim}
\usepackage{ulem}
\usepackage{mciteplus}
\usepackage{amssymb}
\usepackage{amsmath}
\usepackage{graphicx,color}
\usepackage[title]{appendix}
\usepackage{tikz}
\usepackage[all]{xy}
\usepackage{verbatim}
\usepackage{upgreek}
\usetikzlibrary{arrows,shapes}
\def\beq{\begin{equation}}
\def\eeq{\end{equation}}
\def\bea{\begin{eqnarray}}
\def\eea{\end{eqnarray}}

\def\bc{\begin{center}}
\def\ec{\end{center}}

\begin{document}

%\preprint{}

\title{Nucleosome sliding can influence the spreading of histone modifications}


\author{Shantanu Kadam\textsuperscript{1*}, Tripti Bameta\textsuperscript{2*}, Ranjith Padinhateeri}
\email[]{shantanurk123@gmail.com, \textsuperscript{*}tripti.bameta@gmail.com, \textsuperscript{*}ranjithp@iitb.ac.in}
%\thanks{}
\affiliation{Department of Biosciences and Bioengineering, Indian Institute of Technology Bombay, Mumbai, India \\
\textsuperscript{2}Department of Pathology, Tata Memorial Centre, Homi Bhabha National Institute, Mumbai, India}


\date{\today}

\begin{abstract}
Nucleosomes are the fundamental building blocks of chromatin that not only help in the folding of chromatin but also in carrying epigenetic information. It is known that nucleosome sliding is responsible for dynamically organizing chromatin structure and the resulting gene regulation. Since sliding can move two neighboring nucleosomes physically close or away, can it play a role in the spreading of histone modifications ? We investigate this by simulating a stochastic model that couples nucleosome dynamics with the kinetics of histone modifications. We show that the sliding of nucleosomes can affect the modification pattern as well as the time it takes to modify a given region of chromatin. Exploring different nucleosome densities and modification kinetic parameters, we show that nucleosome sliding can be important for creating histone modification domains. Our model predicts that nucleosome density coupled with sliding dynamics can create an asymmetric histone modification profile around regulatory regions. We also compute the probability distribution of modified nucleosomes and relaxation kinetics of modifications. Our predictions are comparable with known experimental results.
\end{abstract}

%\keywords{}

\maketitle

%\section{Introduction}
%\subsection{}
%\subsubsection{}

\section{Introduction}
In cells, DNA is folded and wrapped around octamers of histone proteins forming an array of nucleosomes. Nucleosomes are considered to be the fundamental repeating unit of chromatin and its positioning is important for gene regulation. In typical chromatin, two neighboring nucleosomes are separated by short segments of linker DNA of lengths ranging from 10 to 60 bp~\cite{alberts2002,van2012chromatin,kornberg1974,cortini2016}. Recent advancements in experimental and computational methods have helped us to understand how nucleosomes are organized along DNA\cite{kaplan,widom,segal,milani,zhang,kornberg,ranjith2011,parmar2014,tief2009,van2012sequence,morozov2009,chereji2018,z2019,jiang2021,teif2016p,teif2019p,tripti2018}. 

Nucleosomes also carry epigenetic information in the form of histone modifications apart from the folding of chromatin.
%On each histone protein, there are many covalent chemical modifications and they are also involved in regulating cellular processes including gene expression, DNA repair and DNA replication\cite{erdel2017rev,allis2007,ramachandran2015}.
Specific amino acid residues of histone proteins carry chemical modifications like methylation and acetylations as histone marks\cite{alberts2002,van2012chromatin}. At specified locations on each histone protein, certain enzymes add or remove relevant chemical groups leading to a pattern of  post-translational modifications along chromatin contour\cite{young2005genome,young2011genome}. How these marks get organized along the chromatin is crucial for regulating cellular processes like gene expression, DNA repair, DNA replication etc \cite{erdel2017rev,allis2007,ramachandran2015,reinberg2009,reinbergscience}.
%specific amino acid residues of histone proteins, certain enzymes add or remove certain chemical groups (e.g. methyl, acetyl, phosphoryl) leading to methylation and acetylation of proteins. 
%It is known that histone modifications are dynamic and specific modifying enzymes are recruited by tethering them in a local region along the chromatin.  
%About a century ago spreading of modifications in the context of gene silencing had been observed along a chromosome in irradiated flies\cite{muller1930}. 
Experimentally, one can measure the pattern of histone modification at a given instant in a population of cells by ChIP-Seq methods\cite{zhang2018exp,lindeman}. However, it is a difficult task to measure the modification dynamics in individual cells in real-time\cite{hayashi2009visualizing,hayashi2011tracking,sato2019}. Moreover, comprehensive mechanisms that leads to dynamic histone modification patterns are not fully understood yet.

The phenomena of spreading and subsequent maintenance of histone modifications have been experimentally studied with great interest. Several studies\cite{hall2002,schotta2002,felsenfeld2003nature,shiv} have investigated the formation of heterochromatin and epigenetic inheritance. In these studies, the modified nucleosomes recruit enzymes to similarly modify neighboring unmodified nucleosomes based on a linear stepwise process. Also there are studies, where researchers have tried to unravel how histone modifications spread along chromatin fiber from a given initiation site\cite{allshire2018,allshire2013}. All these studies have considerably contributed to the understanding of modification spreading.

Several theoretical models\cite{cortini2016,erdel2017rev,zhang2015} have been developed to provide insights into the dynamics of histone modifications. Over the years, Sneppen and coworkers have developed models that explain different aspects of histone modification spreading and inheritance~\cite{dodd2007,sneppen2008,sneppen2011,sneppen2016}. They have proposed that long-range interactions lead to a bistable paradigm for a certain range of parameters. 
%At the same time, Sedighi and Sengupta\cite{sengupta2007} proposed a coarse-grained model to study epigenetic silencing in yeast. Here, it was seen that histone deacetylation and binding of Sir protein complexes leads to spreading of silencing. 
Crabtree et al proposed a linear propagation scheme to explain patterns in H3K9me3 that involved localized peaks and soft borders of heterochromatic islands~\cite{hathaway2012dy}. In their model\cite{hathaway2012dy}, they incorporated nucleation, propagation, and turnover rates for modifications, which were necessary to describe H3K9me3 domains. 
In a separate work, they extended the model to estimate several dynamic quantities predicting domain sizes for different values of rates\cite{hodges2012}. This standard model was also extended to incorporate spreading beyond nearest neighbors. There are also other stochastic models~\cite{anink2014m,michieletto2020} that include recruitment, diffusion, long- range interactions leading to the formation of modification patterns. There has been also a Potts-type model by Zhang et al.\cite{zhang2014PRL}, stressing the local nature of interactions, and a model introduced by  Binder et al.\cite{binder2011} investigating epigenetic silencing in eukaryotes.
Another set of models explicitly account for 3D looping and investigate the role of looping in the spreading of histone modifications~\cite{erdel2016,michieletto2016,jost2018NAR, sandholtz2019,sandholtz2020,sandholtz2020p,dave2021}. In these models, the coupling of 3D configurations with spreadings of histone modifications is investigated. 

Nearly all the modelling studies have assumed that nucleosomes are static and modifications can spread to the nearest neighbor nucleosomes independent of the distance between the two nearby nucleosomes. It is plausible that the two neighboring nucleosomes are far away and the spreading may get hindered due to the large gap between those nucleosomes\cite{szerlong2011,bednar1998}.  One way two neighboring nucleosomes can regulate the gap --- internucleosomal distance, or linker length --- is via sliding of nucleosomes\cite{narlikar2013,narlikar2016}. None of the existing models account for the role of the sliding of nucleosomes in the context of modification spreading. In this manuscript, we propose a stochastic model to study the spreading and maintenance of histone modifications taking into account the role of nucleosome sliding.

In the model, we have included  sliding of nucleosomes (due to remodeling complexes), modification of nucleosomes (by modification enzymes), and removal of a modification mark (by de-modifying enzymes) as kinetic events accompanied by their respective rates. We aim to explore whether nucleosome sliding events play any kind of role in spreading the modifications in a particular genomic region, and  how sliding could couple with nucleosome density to determine the modification dynamics.

We have organized this article as follows: first, we have explained the features associated with our one-dimensional model along with simulation details. In the results section, first, we have studied the variation of mean modification spreading times (MMST) for different sliding rates and de-modification rates. In the next section, we have computed probabilities of modified nucleosomes at two different nucleosome densities and compared them with existing experimental results. We have studied the dynamics of modified nucleosomes by estimating their statistical quantities. In the last section, the relaxation dynamics of modified nucleosomes are reported and thoroughly analyzed their behavior. Finally, we have provided conclusions of our work along with some suggestions for new experiments
to test our predictions. 

\section{Model}
%\newpage 
\begin{figure}[h]
\centering
\includegraphics[width=0.85\textwidth]{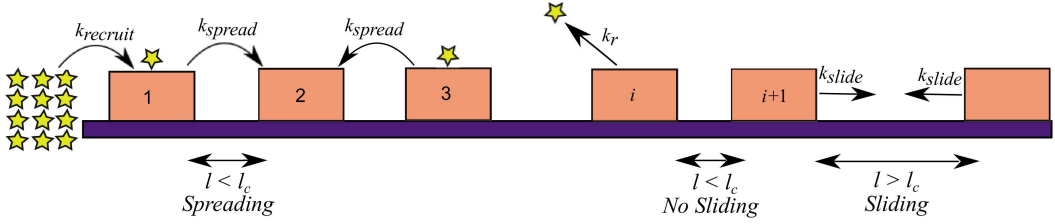}
\caption{{Schematic of the model:}   Histone modifying enzymes (``writers'') are recruited with a rate $k_{recruit}$ at the left boundary. The spreading of the modification (yellow star) occurs with a rate $k_{spread}$, when the distance between two neighboring nucleosomes (brown blocks) is less than $l_{c}$. The de-modification of a nucleosome occurs randomly with a rate $k_{r}$ at any  modified nucleosome. The sliding of a  nucleosome occurs at a rate $k_{slide}$ with a step size of 10 bp, provided that there is a free linker DNA space $\geq l_{c}$ to slide. We take $l_c=10$bp in this work.}
\label{modelf} 
\end{figure}

%\noindent
 
This section describes the model (see Fig.~\ref{modelf}) that we use in our simulations.
We have modelled the spreading of modifications in a quantitative way by simplifying the 3D structure of DNA and focussing on a small section of DNA taking it as a 1D lattice. Thus, in the model, the DNA is considered as a one-dimensional lattice (violet bar) of length $L=5000$ bp. The nucleosomes (brown rectangular blocks) with indices $i=1,2,..., N$ on the DNA are modeled as hard-core particles with each one occupying  $k$ = 147 bp along the lattice.
The hardcore steric interactions among the nucleosomes are modeled by prohibiting a lattice site from getting occupied simultaneously by more than one nucleosome. 

In the model, we have considered four kinetic events: i) recruitment of modification enzymes with the rate $k_{recruit}$, ii) transfer of this modification enzyme to an unmodified nucleosome with rate $k_{spread}$, iii) fall of modification enzyme with the rate $k_{r}$ and iv) sliding of nucleosomes along the DNA with the rate $k_{slide}$. In this study, the binding and dissociation of nucleosomes are ignored. Hence, we have assumed that the total number of nucleosomes ($N$) is constant.

We consider a situation where modification enzymes are recruited at a specific location {$i=0$} (left end of the lattice) with a rate $k_{recruit}$ (Fig.~\ref{modelf}). This is implemented when the nucleosome ($i=1$) gets closer to the source of modification enzymes within a distance (gap between two nucleosomes) less than $l_{c}$ = 10 bp. The recruited modification enzyme further spreads along the lattice with the rate $k_{spread}$ to a neighboring unmodified nucleosome provided that inter-nucleosomal distance is less than  $l_{c}$ = 10 bp. One of the hallmarks of histone modification spreading is the positive feedback\cite{dodd2007,erdel2016}. In this model a modified nucleosome inducing modification to spatially close ($\leq$ 10bp) neighbor is essentially the positive feedback.
In the simulations, it is assumed that the rate of recruitment ($k_{recruit}$) is the same as the rate of spreading ($k_{spread}$). The modified nucleosome can be randomly demodified by removal of modification enzyme with a rate $k_{r}$. Our simulations aim to answer the following question, how nucleosome sliding affects the spreading of histone modifications ? Hence, we have incorporated random nucleosome sliding to the left or right with a rate $k_{slide}$ per nucleosome. This rate represents the rate of reaction by ATP-dependent chromatin remodelers\cite{petersson2014} that are responsible for such repositioning of nucleosomes in cells. It is assumed that, sliding step size is 10 bp such that the diffusion constant is $\sim k_{slide} (10 {\rm bp})^2$. In our studies, it is assumed that beyond $i=1$ and $i=N$ there are boundary elements, which do not allow the spreading any further. 

We have performed kinetic Monte Carlo simulations using Doob-Gillespie Algorithm\cite{doob42,doob45,dan76,dan77} using the rates of sliding, modification, and de-modification events. All these events are independent of each other. In this way, the simulation was run for some desired time. The nucleosome density is calculated as, $\rho = \frac{147*N}{L}$; where $N$ is the total number of nucleosomes and $L$ is the total length of the 1D lattice. In all the simulations discussed in this work, the modification rate, $k_{spread}$ is kept fixed at 1 $s^{-1}$.  The rates ($k_{slide}$, $k_{r}$) of all other events are scaled with $k_{spread}$  giving dimensionless quantities for those respective rates. The time reported in the simulations is taken in the unit $\tau_{s} = \frac{1}{k_{spread}}$. All the simulations in this study were performed by taking an average over 2000 independent runs.

\section{Results}
\subsection{Kinetics of modification spreading}
\subsubsection{Mean Modification Spreading Time: Simulations and Mean Field Theory}
We simulated the spreading of histone modification as discussed in the Model section for various nucleosome densities  computing mean modification spreading time (MMST) as a function of sliding rates. The MMST is defined as the time required for the first successful modification of the last nucleosome $(i=N)$, given that the modification spreads from the initiation site $(i=0)$. These are similar to the mean first passage time calculations in statistical mechanics\cite{redner2001},  which are a measure of modification spreading time. 

\begin{figure}[h!]
\centering
\includegraphics[width=0.70\textwidth]{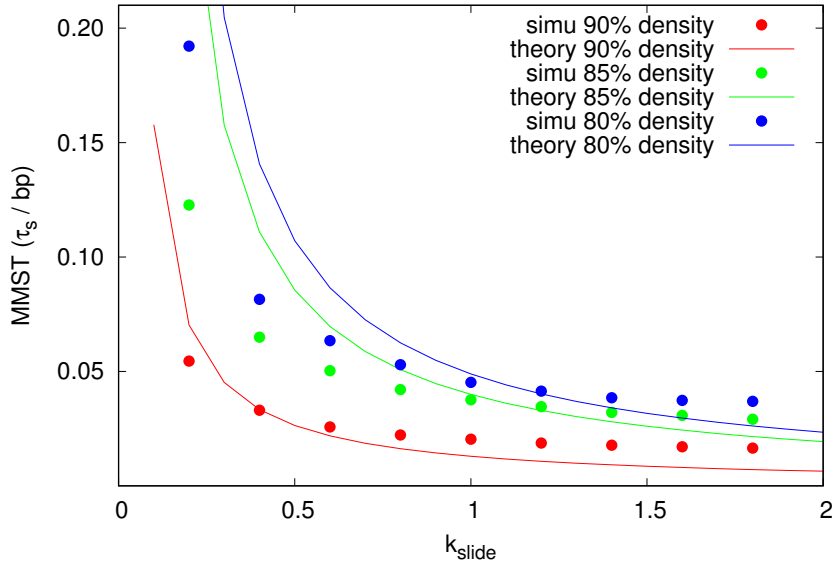}
\caption{{Quicker sliding reduces the spreading time:} The mean modification spreading time (MMST) with $k_{slide}$ using simulations (dots) and mean-field theoretical calculations (lines) for different nucleosome densities, viz. : 90\% (red), 85\% (green) and 80\% (blue). All rates are measured in units of $k_{spread}$.} 
\label{mmst4}
\end{figure}

We observed that, as the sliding rate increases, the mean modification spreading time decreases and saturates to a constant value(Fig.~\ref{mmst4}). For a zero sliding rate, on a sufficiently long DNA, the probability of finding at least one pair of nucleosome-neighbors with a gap (linker length) greater than 10 bp is very high. Hence, for zero sliding rate, the modification may not reach the other end implying the mean modification spreading time can be infinity. In order to see the effect of sliding, the mean modification spreading time (MMST) was computed at three different nucleosome densities (90\%, 85\%, and 80\%). Here, the de-modification rate, $k_{r}$, of nucleosome was kept fixed at 0.01.

%\begin{table}[!ht]
%\centering
%\caption{{\bf Parameter values used in the paper.}}
%\begin{tabular}{c|c|l|l}
%\hline
%\bf{Parameter} & \bf{Description} & \bf{Value} & \bf{Ref.} \\ \hline
%$L$ & Length of the 1D lattice & 5000 bp  & \\ \hline
%$k$ & Nucleosome size & 147 bp &  \\ \hline 
%$^a$$k_{slide}$ & sliding rate of nucleosome &0.25 - 2 & [12] \\ \hline 
%$^a$$k_{r}$ & de-modification rate of nucleosome &0.01 - 0.1 & [31] \\ \hline
%\end{tabular}
%\label{table1} \\
%\footnotesize{$^a$$k_{slide}$ and $k_{r}$ are in units of $k_{spread}$}
%\end{table}

We also used a  mean-field theoretical study to understand how MMST would vary with sliding rates for different densities. For this calculation, at a given density, we have taken nucleosomes to be homogeneously distributed along the lattice. The effect of nucleosome sliding was incorporated in the effective spreading rate of modification, $k_{se}$,  which is a function of the sliding rate of the nucleosome and average inter-nucleosomal distance. 
%\textcolor{red}{Note that we have approximated the statistical distribution of nucleosomes to be distributed homogeneously for a given density.} 
The mean modification spreading time ($T_{i\rightarrow n}$), from $i^{th}$ to $n^{th}$ nucleosome follows the difference equation\cite{gardiner}: 

\begin{equation}
T_{i\to n}=\dfrac{1}{k_{\rm se}+k_{r}}+\dfrac{k_{se}}{k_{se}+k_{r}}T_{i+1\to n}+\dfrac{k_{r}}{k_{se}+k_{r}} T_{i-1\to n}
\end{equation}
where index $i$ varies from $0$ to $N$. The index $i=0$ represent nucleation site, while ({$1$}) and ($N$) are indices of first and last nucleosome respectively. The $N$ is the total number of nucleosomes in the model. From solving this $N+1$ set of linear equations (see Appendix) :
\begin{equation} 
T_{0\to N}=  \frac{1}{{ (k_{\rm se})^N} }\sum_{\ell=1}^{N}  (N-\ell +1 )  (k_{\rm se})^{N-\ell} ~~k_{\rm r}^{\ell-1} 
\label{T1n}
\end{equation}%\eea
It is expected, when de-modification rate of nucleosomes is set to zero (i.e. $k_{\rm r}=0$) (2) reduces to $T_{0\to N}=\frac{N}{k_{\rm se}}$. In this equation, $k_{\rm se}$ depends on the sliding rate of nucleosome and the inter-nucleosomal distance via the relation\cite{balakrishnan1,balakrishnan2}: 
\begin{equation}
k_{se} = \frac{l_s^2 \times k_{slide}}{gap^2}
\end{equation}
where $l_s=10$ bp is the step size of sliding events, and the gap is the linker length. This relation can be understood as an inverse of the time scale of the meeting of two nucleosomes. In Fig.~\ref{mmst4}, we plot the mean-field theory results (Eq.\ref{T1n}, curves) along with simulation results (dots). For certain nucleosome densities, both results are comparable. A significant variation was observed in MMST when density was varied from 90\% to 85\%. The higher values of MMST at lower densities were a signature of the presence of long gaps (greater than 10 bp) between nucleosomes, which eventually slowdowns the spreading of modifications. This also implies, when $k_{slide} \ll k_{spread}$ MMST is higher, while in the opposite limit it is found to be smaller. 

We have got elevated profiles of MMST at 85\% and 80\% nucleosome densities implying reduced densities contributing towards an increase in MMST values. For low sliding rates (less than or around 0.5), the MMST changes a lot with the sliding rate indicating that the longer times it takes to spread the modifications across the lattice. For higher sliding rates, MMST change is small. Altogether, all these results imply that the sliding of nucleosomes can play a significant role in spreading the modification across the lattice.

\subsubsection{De-modification Events Increase Spreading Time} 
\begin{figure}[h!]
\centering
\includegraphics[width=0.70\textwidth]{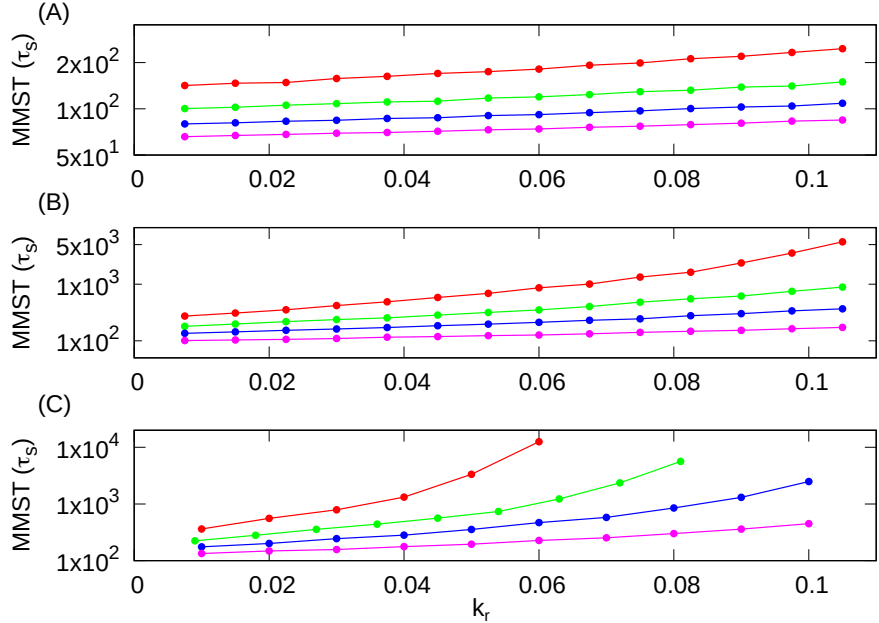}
\caption{{Spreading time increases with de-modification events:} The mean modification spreading time (MMST) as a function of de-modification rate $k_{r}$ for different nucleosome densities : (A) 90\%, (B) 85\% and (C) 80\%. Different curves are for sliding rates 0.5, 1, 2, and 6 from top (red) to bottom (magenta). All rates are measured in units of $k_{spread}$.} 
\label{kremove} 
\end{figure}

In reality, modified nucleosomes can get de-modified and such de-modification events may be 
crucial in some contexts. To examine the effect of de-modification on MMST, we varied the de-modification rate ($k_r$) over a range. In Fig.~\ref{kremove} we present our results for MMST  by changing the de-modification rates of nucleosomes for four different sliding rates 0.5 (red), 1 (green), 2 (blue) and 6 (magenta). Since, the residence time of the modification\cite{anink2014m,phair2004} is more than spreading time, we take de-modification rate  $k_{r} \ll 1$ in units of $\frac{1}{\tau_{s}}$. 

%\newpage

In Figs.~\ref{kremove} (A), (B) and (C) MMST results are plotted for different nucleosome densities 90\%, 85\%, and 80\% respectively. Here, one sees an interplay between nucleosome de-modification rate and sliding rate. An increase in de-modification rates contributes to the corresponding increase in the MMST. It would take a longer time for spreading the modification across the lattice for larger values of de-modification rates. However, an increase in sliding rates from 0.5 to 6 (red, green, blue, magenta curves) has had contributed to a substantial decrease in the MMST. It was found that with a reduction in nucleosome density, the gaps between nucleosomes as well as de-modification rates have contributed to an increase in modification spreading times.

\subsection{Domains of modified nucleosomes : effect of sliding and de-modification events} 
In this section, we discuss how the modified domains of nucleosomes are maintained for different sliding and de-modification rates at a fixed nucleosome density. We have calculated the probability of modified nucleosomes at a steady state. The simulations were carried out by varying sliding rates whilst fixing $k_{r}$ = 0.1. 

\begin{figure}[h]
\centering
\includegraphics[width=0.80\textwidth]{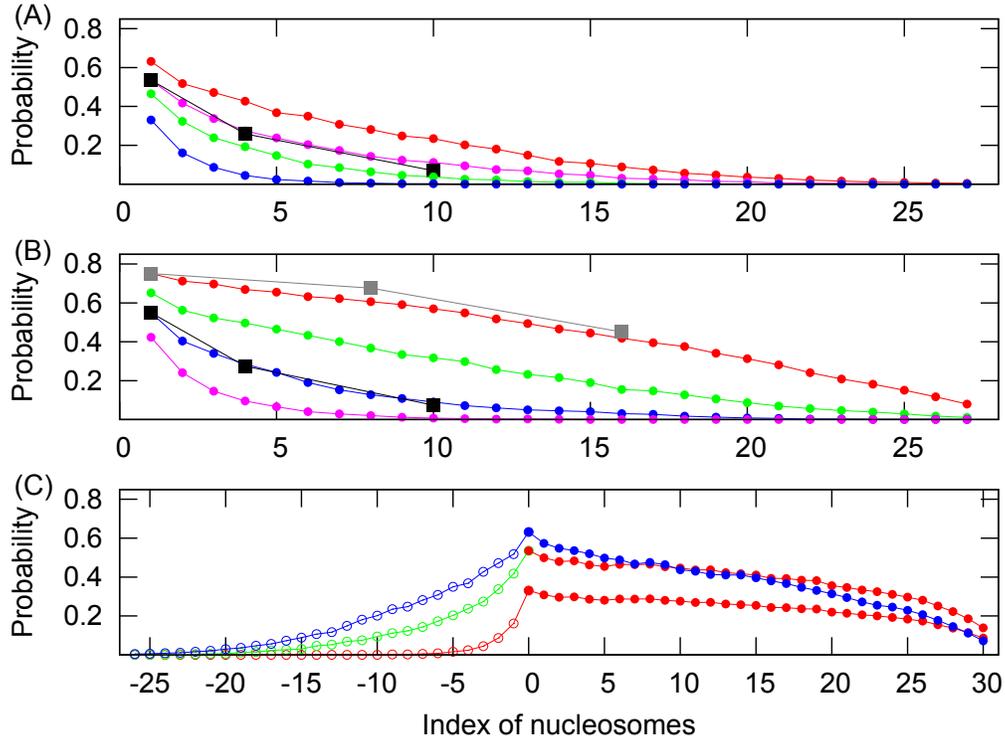}
\caption{{Probabilities with sliding and de-modification rates:} The probability of modified nucleosomes for sliding and de-modification rates at 80\% nucleosome density: (A) for $k_{r}$ = 0.1, $k_{slide}$ = 2.2 (red), 1 (magenta), 0.5 (green) and 0.1 (blue) and (B) for $k_{slide}$ = 0.25, $k_{r}$ = 0.02 (red), 0.03 (green), 0.05 (blue), 0.1 (magenta). The black squares depict experimental data (Hathaway et al. (\cite{hathaway2012dy})) for MEF cells and grey squares for ES cells.(C) Probabilities at 80\% density (left of initiation site) for $k_{slide}$ = 0.1 (red), 1 (green) and 2.2 (blue) open circles and 90\% density (right of initiation site) for $k_{slide}$ = 0.1 (red) and 0.075 (blue) filled circles. All times are measured in units of $\tau_{s}$.} 
\label{lastfig} 
\end{figure} 

The probability of finding modified nucleosomes at any location along the DNA contour, at 80\% nucleosome density, is depicted in Fig.~\ref{lastfig} with different colors representing different sliding and de-modification rates. The probability sharply decays as we decrease the sliding rate (Fig.~\ref{lastfig}(A)).  In other words, the modification pattern tends to get more localized with a peak at the initiation (source) site. This sharp decay in modification profile is similar to what is observed in experiments near the nucleation site. For  $k_{slide}$ = 1, our simulation results are comparable to 
experimental H3K9me3 ChIP data from Hathaway et al.\cite{hathaway2012dy} of mouse embryonic 
fibroblasts (MEFs).  A similar set of simulations were repeated at 90\% nucleosome density for different sliding rates (see Fig S4-A). In this case, density has played an important role in modification of these nucleosomes along with their sliding rates. A significant decrease in probabilities at the end of lattice is a signature of de-modification events dominating over the spreading of modifications. 
    
As a next step, simulations were carried out by varying de-modification rates while fixing the $k_{slide}$ = 0.25 (Fig.~\ref{lastfig}(B)). For small $k_{r}$ values, like 0.02 and 0.03, the modification pattern has a larger spread. It was found that for  $k_{r}$ = 0.02, our simulation results are comparable to experimental H3K9me3 ChIP data from Hathaway et al.\cite{hathaway2012dy} of  Embryonic Stem (ES) cells. For relatively higher $k_{r}$ values localized modification pattern is observed with a peak near the initiation site. At $k_{r}$ = 0.05 the simulated modification profile is comparable to what is observed in experiments for MEF cells. 

Our results so far suggest that combination of different sliding rates and nucleosome density can lead to very different modification patterns. Since it is plausible that on either side of certain boundary or boundary elements (e.g. transcription start site), nucleosome density and action of chromatin remodellers could be very different. We explore this to examine whether this can lead to an asymmetry in nucleosome modification spread patterns\cite{sandholtz2020}. We simulated nucleosomes with different densities and sliding rates on either side of a boundary (initiation site). The results are in Fig.~\ref{lastfig}(C), where the asymmetry in the positioning of modified nucleosomes about the initiation site can be seen. The density of the nucleosomes are always fixed at 90\% on the right-hand side of the initiation site and at 80\% on the left-hand side. Then different combinations of sliding rates are taken on both sides of initiation site. The top blue curve (with open circles on the left of initiation site for $k_{slide}$ = 2.2 and filled circles on right of initiation site for $k_{slide}$ = 0.075) has least asymmetry; the bottom red curve (with open circles on the left of initiation site and filled circles on right of initiation site for $k_{slide}$ = 0.1) has maximum asymmetry. Thus, we suggest that variability in sliding rate and nucleosome densities is a potential way of explaining the asymmetry across various boundary elements.

These results together show that the interplay between nucleosome de-modification rate and the 
sliding rate determines the profile of the modification pattern. At low sliding and high de-modification rates, the modification is peaked near the source and decays quickly. For high sliding rate and low de-modification rate, the decay is more gradual. 

\subsection{Dynamics of modified nucleosomes : estimation of statistical quantities} 
\begin{figure}[h]
\centering
\includegraphics[width=0.65\textwidth]{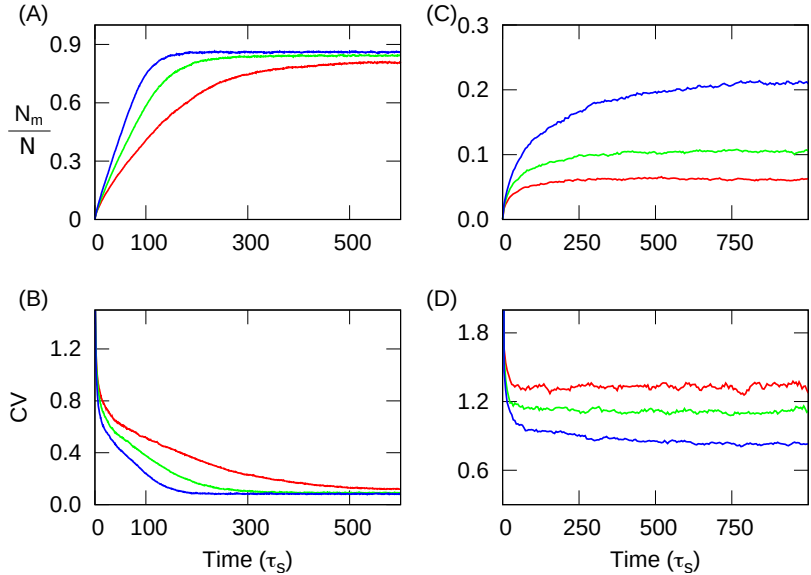}
\caption{{Statistics of modified nucleosomes:} The (A) fraction of modified nucleosomes and (B) coefficient of variation (CV) of modified nucleosomes for 90\% nucleosome density (N = 31 nucleosomes) (C) fraction of modified nucleosomes and (D) coefficient of variation (CV) of modified nucleosomes for 80\% nucleosome density (N = 27 nucleosomes). All plots are for different nucleosome sliding rates $k_{slide}$ =  0.5 (red), 1 (green) and 2 (blue) for lattice length $L$ = 5000 bp. All times are measured in units of $\tau_{s}$.} 
\label{ninety} 
\end{figure}

In this subsection, we discuss the time evolutions in the amount of modification and its fluctuations at biologically relevant nucleosome densities (80\% and 90\%). For each time step, we plot the mean fraction of modified nucleosomes $\frac{\rm{N_{\rm{m}}}}{\rm{N}}$, where  $\rm{N_{\rm{m}}}$ is the mean number of modified nucleosomes (Fig.~\ref{ninety}). 
The fluctuations were quantified using the coefficient of variation (CV), a dimensionless quantity defined as the ratio of standard deviation to mean. 
These quantities were calculated for different sliding rates $k_{slide}$ = 0.5, 1, 2.
%; while their probability distributions for $k_{slide}$ = 1 $s^{-1}$ and 5  $s^{-1}$. 
$k_{r}$ was kept fixed at 0.1. At  $t=0$, $\rm{N_{\rm{m}}}$ is taken as zero. As expected, $\frac{\rm{N_{\rm{m}}}}{\rm{N}}$ increases with time and saturates. Here different trajectories (colors) correspond to different rates of sliding. At higher density, the variation among $\frac{\rm{N_{\rm{m}}}}{\rm{N}}$ for different sliding rates (i.e. different curves with different colors) is small. However, at lower density (80\%), the mean dynamics show huge variation as we change the sliding rates, suggesting that steady-state as well as the dynamics at low densities are crucially affected by the rates. 

\begin{figure}[h]
 \centering
  \includegraphics[width=0.65\textwidth]{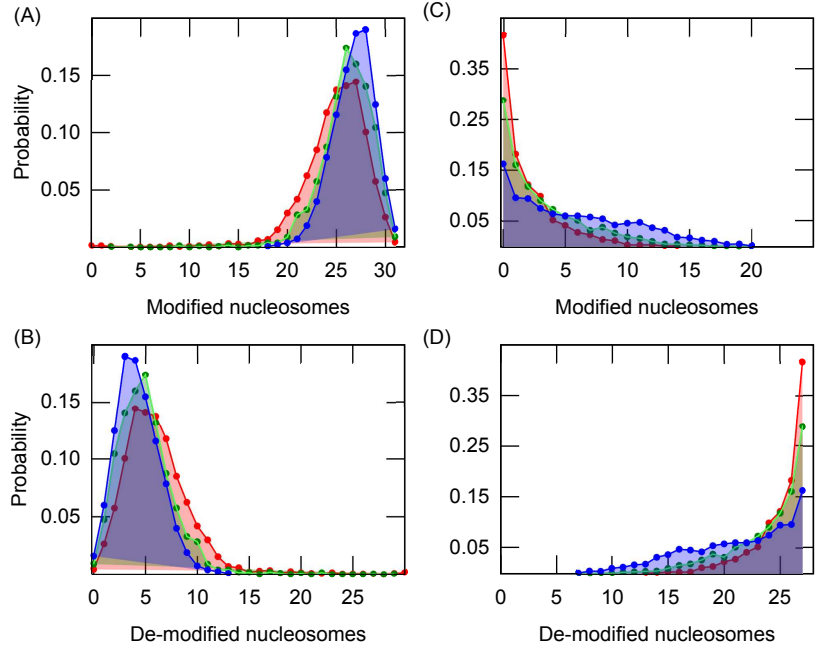}
 \caption{{Number distribution profiles of modified and de-modified nucleosomes:} The probability distributions of number of (A) modified and (B) de-modified nucleosomes at 90\% nucleosome density. The probability distributions of number of (C) modified and (D) de-modified nucleosomes at 80\% nucleosome density for nucleosome sliding rates $k_{slide}$ = 0.5 (red), 1 (green) and 2 (blue). All rates are measured in units of $k_{spread}$.} 
\label{dist90}    
\end{figure} 

We find that CV values are decreasing as a function of time. This could be because at early times the modification numbers are less and hence the fluctuation is high.  At 90\% density, CV is smaller than unity (standard deviation smaller than the mean) and CV is similar at the steady-state for all sliding rates. However, for 80\% nucleosome density, the CVs show well-separated steady states, and fluctuations are comparable or bigger than the mean (CV is comparable or above $1$). We found almost the same steady states for fraction of modified nucleosomes and CV, when the simulation was done at 80\% nucleosome density with all nucleosomes modified (N = 27) at t = 0. It implies that the final states are independent of the initial conditions. (whether all nucleosomes or none of them were modified) (see Fig S1)  
%At 90\% density, the spreading time is of the order of 300 s, which is less than cell cycle time. At lower density spreading time is high implying we need higher densities and higher sliding rates for quick spreading. It also implies lower densities and lower sliding rates can stop spreading by creating boundaries.     

The probability distributions for the number of modified and de-modified nucleosomes at 90\% and 80\%  nucleosome densities are depicted in Fig.~\ref{dist90}. These distributions were obtained by finding nucleosome numbers at steady states. In Fig.~\ref{dist90}(A), for given sliding rates distributions of modified nucleosomes are negatively skewed. It was found that at 90\% density almost all nucleosomes have got modified for lower (0.5)  and higher (2) values of sliding rates. This implies at high nucleosome density modification is not much affected by sliding rates of nucleosomes. The distribution of de-modified nucleosomes is shown in Fig.~\ref{dist90}(B). At 80\% nucleosome density, in Fig.~\ref{dist90}(C and D) qualitatively same distributions were observed for low and high values of sliding rates. At a lower sliding rate (0.5) less number of nucleosomes were modified; while at a higher sliding rate modification spreads along the lattice modifying a large number of nucleosomes. 

\subsection{Relaxation dynamics of modifications when the initiation site is removed} 
   
\begin{figure}[h]
\centering
\includegraphics[width=0.65\textwidth]{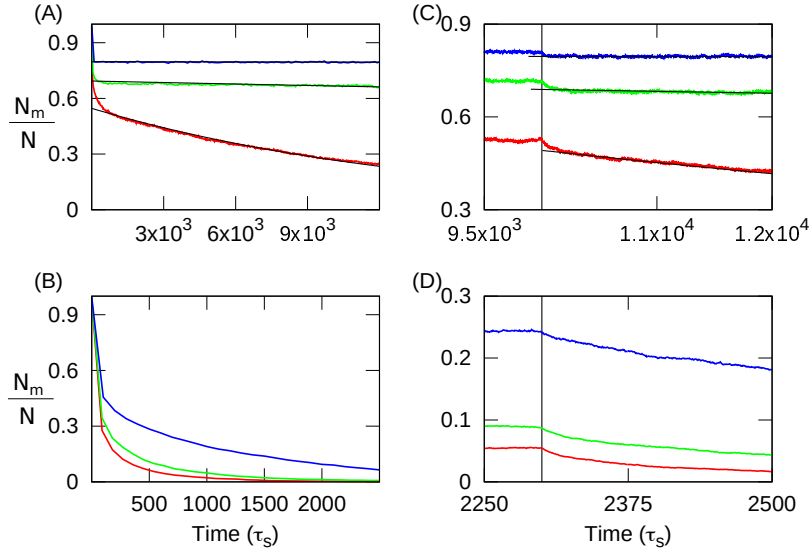}
\caption{{Relaxation of modified nucleosomes for different sliding rates :} The time profiles of fraction of modified nucleosomes at nucleosome densities (A) 90\% (N = 31 nucleosomes) and (B) 85\% (N = 29 nucleosomes) when all nucleosomes are modified at t = 0 and $k_{recruit}=0$. 
At nucleosome densities (C) 90\% and (D) 85\% at t = 0, $k_{recruit}=1$ and all nucleosomes are unmodified. At steady state (black vertical line) of modification we take $k_{recruit}=0$. All the results are for sliding rates $k_{slide}$ = 0.1 (red), 0.2 (green), 0.5 (blue). The fitted curve is shown in black and all times are measured in units of $\tau_{s}$.} 
\label{rela90}    
\end{figure} 
%\newpage
The cell actively maintains the average number of modified nucleosomes by constantly inserting modifications. In this subsection, we will examine how the number of modified nucleosomes decreases as a function of time if the nucleation site removed (spreading from the nucleation site is switched off). We will discuss two different cases.  In the first case, at t = 0 all nucleosomes are assumed to be in the modified state. As the time starts, we stop influx of histone modification enzymes from the initiation site ($k_{recruit}=0$). That is, starting from a fully modified state, we fix $k_{recruit}=0$ and simulate dynamics taking all other events. In the second case, we simulate the full system to obtain the steady state, and at some time point in steady state, we set $k_{recruit}=0$.

In Fig.~\ref{rela90}(A) we present the results for the first case with higher density (90\%) and different  nucleosome sliding rates (0.2 (green), 0.5 (blue) and (0.1 (red)). We observe stable patterns of modified nucleosomes are maintained for sliding rates 0.2 and 0.5 in respective steady states. However, when the sliding rate is low (0.1 (red)) the fraction of modified nucleosomes decreases as a function of time. This curve was fitted with an exponentially decaying function giving rise to a decay rate of the order of $10^{-6}$ per unit time. This is analogous to the effective de-modification rate in our simulations. The loss of steady-state pattern emphasizes the importance of sliding events even at such a higher density. However, at lower densities in (85\%) Fig.~\ref{rela90}(B), such steady-state patterns are absent for all the sliding rates simulated. However, for higher sliding rate, the mean modification decays slowly. 

In the second case, we simulated the full system (taking all events discussed in subsection C) until the steady state. At a particular time point indicated by the vertical bar in Fig.~\ref{rela90}(C) and (D), influx of modification was switched off ($k_{recruit}$ is set to $0$). For higher sliding rates  (0.2 (green) and 0.5 (blue)) and higher density, there is no dip in the fraction of modified nucleosomes. However, for a lower sliding rate  (0.1(red)) the fraction of modified nucleosomes decreases a bit but stays near the steady state. At 85\% nucleosome density, the fraction of modified nucleosomes decay  by slightly but stays closer to their respective steady states. Fitting the decay curve with an exponential function, the decay constant at  85\% nucleosome density is found to be around $10^{-3}$ per unit time (see Supplementary Table S1 and Table S2). 

In short, the interplay between density and sliding decide the relaxation dynamics of modifications;  an increase in $k_{slide}$ values (from 0.1 to 0.5) the modified nucleosomes decay slowly. The sliding of the nucleosome is contributing to slowing down the decay (also see Supporting information Fig S2 and Fig S3).

\section{Conclusions and suggestions for experiments} 
In this article, we have discussed how nucleosome sliding may affect the spreading of histone modifications along the lattice.
The spread of modifications is quantified by estimating the mean modification spreading time (MMST). It was found that for low nucleosomal sliding rates it takes a longer time for spreading the modification across the lattice; while it took less time for higher sliding rates. We have confirmed these findings by doing an analytical estimation of MMST using a mean field theory. It was also found that the 
interplay between nucleosome sliding events and nucleosome density determines spreading times. The larger de-modification rates have contributed to enhanced modification spreading times, but the sliding rates of nucleosomes have helped to restrict them. The dynamics of modified nucleosomes were studied by computing statistical quantities like fluctuations and probability distributions, which were found to be dependent upon nucleosome densities. We show that for certain densities and sliding rates, the nucleosome modification pattern is localised in a region, as seen in experiments. 
 This work also shows that certain parameters can give raise to asymmetric nature of nucleosome modifications about the initiation site. The interplay between sliding events and density of nucleosomes also influence the relaxation dynamics of modifications. Overall, the proposed model gives insights into the role of sliding events and how the interplay between density and sliding can be an important determinant.  

%\subsection{Suggestions for experiments}
The main prediction of the paper is how the sliding of nucleosomes and nucleosome densities can influence the spreading of modifications. This can be tested by developing appropriate mutants of nucleosome sliding enzymes and examining whether the mutations affect the spreading of modifications or not. One may also design different chromatin arrays having very different nucleosome densities (nucleosome repeat lengths) and examine how these would influence the spreading of modifications. 

\section*{Acknowledgments}
This research has been supported by funds to SK from Department of Science and Technology, India under Science and Engineering Research Board, National Post-doctoral Fellowship (NPDF) with file number: PDF/2017/002502 and to RP from Department of Biotechnology, India Grant BT/HRD/NBA/39/12/2018-19.
%\section*{Supporting information}

%\paragraph*{S1 Appendix}
\begin{appendix}
%\appendix*
%\section{APPENDIX}
%\section{}
\section{Recursive relation of mean first passage time from survival probability}
\label{app:A}
%\label{S1_Appendix} 
A simplified mean field calculation of this model can give a closed form expression for MMST as a function of rates. For this calculation, we take nucleosomes homogeneously distributed along the lattice. The modification spreads from the nucleation site with an effective rate $k_{se}$, which depends on sliding rate and inter-nucleosomal distance by the following relation\cite{balakrishnan1,balakrishnan2}:
\bea
%\begin{equation}
k_{se} = \frac{l_s^2 \times k_{slide}}{gap^2}
%\end{equation}
\eea

For MMST calculations, we first calculate survival probability ($S_{i\to n}(t)$) which is defined as: the probability that modification has not reached to the $n^{th}$ nucleosome till time $t$ given that at $t = 0$  
$i^{th}$ nucleosome was already modified\cite{gardiner}.
\bea
\dfrac{ \partial S_{i\to n}(t) } {\partial t} &=& k_{se} S_{i+1\to n}(t) + k_r S_{i-1\to n}-\left( k_{se}+k_r \right) S_{i\to n}(t).
\eea
We can write first passage time $T_{i\to n}$ as;
\bea
T_{i\to n}&=&\int_{0}^{\infty} t ~F_{i\to n} (t) dt\nonumber\\
&=&\int_{0}^{\infty} t \dfrac{ -\partial S_{i\to n}(t) } {\partial t} dt\nonumber\\
&=&- t~ S_{i\to n}(t)|_{0}^{\infty}-\int_{0}^{\infty} 1. \left(-S_{i\to n}(t)\right) dt \\
&=&\int_{0}^{\infty}  S_{i\to n}(t)  dt 
\eea
First term in Eq.(A3) vanishes because for any bounded survival probability at large time is $\approx 0$, and it approaches zero much faster than $1/t$. 
\bea
\int_{0}^{\infty} \dfrac{ \partial S_{i\to n}(t) } {\partial t}  dt &=&  k_{se} \int_{0}^{\infty} S_{i+1\to n}(t) dt + k_r  \int_{0}^{\infty} S_{i-1\to n} dt -\left( k_{se}+k_r \right)  \int_{0}^{\infty} S_{i\to n}(t) dt\nonumber\\
S_{i\to n}(\infty) - S_{i\to n}(0)&=& k_{se} T_{i+1\to n} + k_r T_{i-1\to n}-\left( k_{se}+k_r \right) T_{i\to n}\\
T_{i\to n}&=&\dfrac{1}{k_{se}+k_r}+\dfrac{k_r}{k_{se}+k_r}T_{i-1\to n}+\dfrac{k_{se}}{k_{se}+k_r} T_{i+1\to n}
\eea 
In Eq.(A5), we have used two properties of survival probability that a system should survive with probability $1$ at $t=0$ and modification should survive with probability $\approx 0$ at very large time. 

Now using Eq.(A6), we can write a recursive equation between MMSTs :  ($T_{i\to N} \rightarrow {\rm MMST~ from}~ {i}^{\rm th}~{\rm nucleosome ~to~} {N}^{\rm th}~{\rm nucleosome}$)

%\section*{S1 Appendix }
	%\bea
 	%\begin{equation*}
	%T_{i\to n} \rightarrow {\rm MFPT~ from~ i}^{\rm th}~{\rm nucleosome ~to~ n}^{\rm th}~{\rm nucleosome}\nonumber
 	%T_{i\to N} \rightarrow {\rm MMST~ from}~ {i}^{\rm th}~{\rm nucleosome ~to~} {N}^{\rm th}~{\rm nucleosome}\nonumber
	%\eea	
	\bea
 	T_{0\to N}&=&\dfrac{1}{k_{recruit}}+T_{1\to N}\nonumber\\
 	T_{1\to N}&=&\dfrac{1}{k_{se}+k_{r}}+\dfrac{k_{se}}{k_{se}+k_{r}}T_{2\to N}+\dfrac{k_{r}}{k_{se}+k_{r}} T_{0\to N}\nonumber\\
 	T_{2\to N}&=&\dfrac{1}{k_{se}+k_{r}}+\dfrac{k_{se}}{k_{se}+k_{r}}T_{3\to N}+\dfrac{k_{r}}{k_{se}+k_{r}} T_{1\to N}\nonumber\\
 	&.&\nonumber\\
 	&.&\nonumber\\
 	&.&\nonumber\\
 	T_{i\to N}&=&\dfrac{1}{k_{se}+k_{r}}+\dfrac{k_{se}}{k_{se}+k_{r}}T_{i+1\to N}+\dfrac{k_{r}}{k_{se}+k_{r}} T_{i-1\to N}\nonumber\\
 	&.&\nonumber\\
 	&.&\nonumber\\
 	&.&\nonumber\\
 	T_{N-1\to N}&=&\dfrac{1}{k_{se}+k_{r}}+\dfrac{k_{r}}{k_{se}+k_{r}} T_{N-2\to N}\nonumber\\
 	T_{N\to N}&=&0
 	\eea

	We can write these equations in matrix form:
 	\bea
 	\mathbb{M}1~ \pmb{T}=\pmb{B} \nonumber
 	\eea
 	\begin{gather}
 	\begin{bmatrix} k_{\rm recruit} & -k_{\rm recruit}& &  &0 \\  -k_{\rm r} & k_{\rm se}+k_{\rm r}&-k_{\rm se}&\ddots&\ddots \\ 0  & \ddots& \ddots & \ddots&\ddots \\    & &&& \\    &\ddots&-k_{\rm r} & k_{\rm se}+k_{\rm r}&-k_{\rm se} \\  0  &&& -k_{\rm r} & k_{\rm se}+k_{\rm r}\end{bmatrix} 
 	\begin{bmatrix} T_{0\to N}\\T_{1\to N}\\. \\ .\\.\\ .\\T_{N-2\to N}\\T_{N-1\to N}\end{bmatrix}
 		=
 	\begin{bmatrix}
 	1\\1\\.\\.\\.\\.\\.\\1
 	\end{bmatrix}
 	\end{gather}
 	$k_{\rm recruit}$ and $k_{\rm se}$ are function of  nucleosome density and  $ k_{\rm slide}$
 	We solve the above equation by taking inverse of matrix $\mathbb{M}$ to get $T_{1\to N}$. For simplicity first  take $k_{\rm recruit} =  k_{\rm se}$ 
 	
 	\bea
 	T_{0\to N}=  \dfrac{1}{{( k_{\rm se})^N} }\sum_{\ell=1}^{N}  (N-\ell +1 )  k_{\rm se}^{N-\ell} ~~k_{\rm r}^{\ell-1} 
 	\eea
 		
%	We can try fitting  $k_{se} \propto \dfrac{k_{slide}}{gap^2} $ to the data. 
\end{appendix}

\end{document}

\section*{Supplementary Information}

\paragraph*{Fig S1}
\label{S1_Fig}
\begin{figure}[h]
\centering
\includegraphics[width=0.60\textwidth]{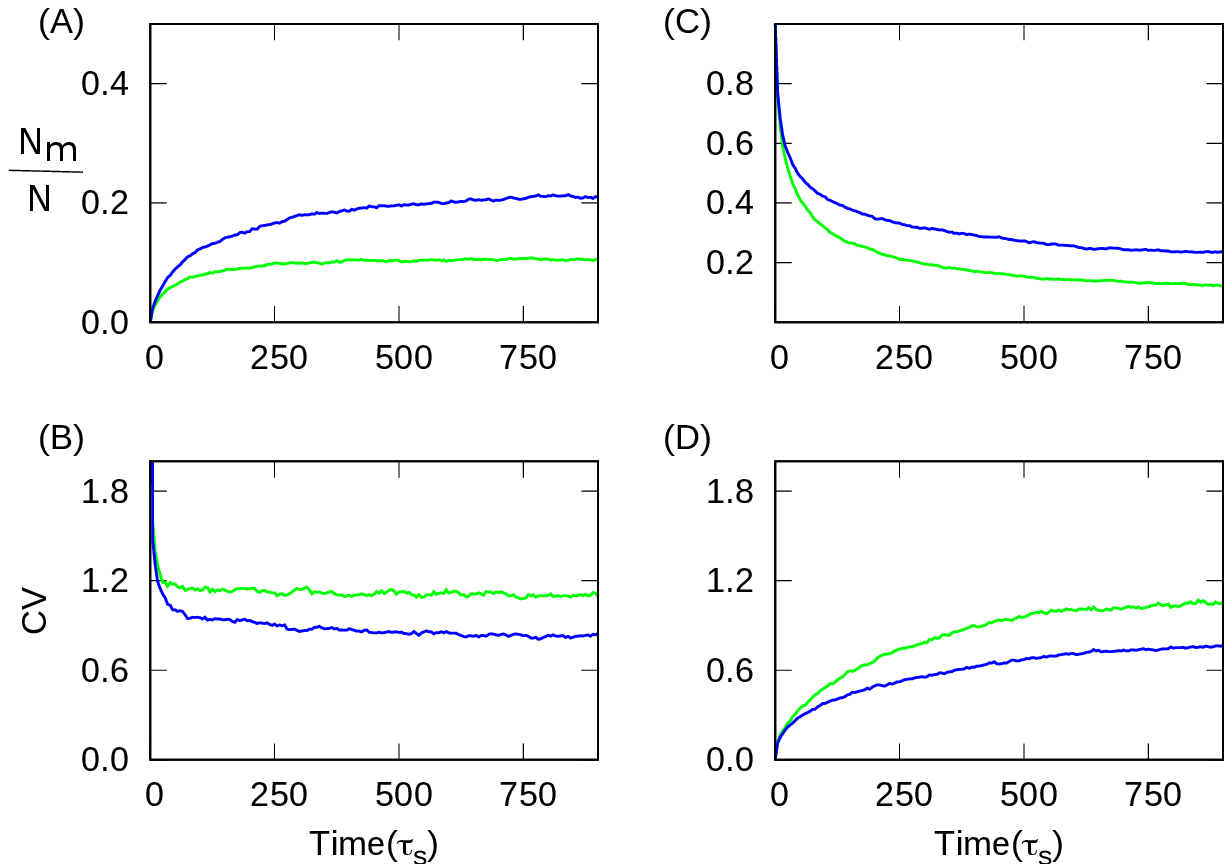} 
\end{figure} 
{\ Similar steady states for different initial conditions.} (A) normalized mean modified nucleosomes and (B) coefficient of variation (CV) when ${\rm{N_{\rm{m}}}}$ = 0 at t = 0, while (C) normalized mean modified nucleosomes and (D) coefficient of variation (CV) when ${\rm{N_{\rm{m}}}}$ = N at t = 0 for 80\% nucleosome density (N = 27) at different nucleosome sliding rates, ($k_{slide}$), 1 (green) and 2 (blue) for lattice length $L$ = 5000 bp. All the times are measured in units of  $\tau_{s}$.

\paragraph*{Fig S2}
\label{S2_Fig}
\begin{figure}[h]
\centering
\includegraphics[width=0.60\textwidth]{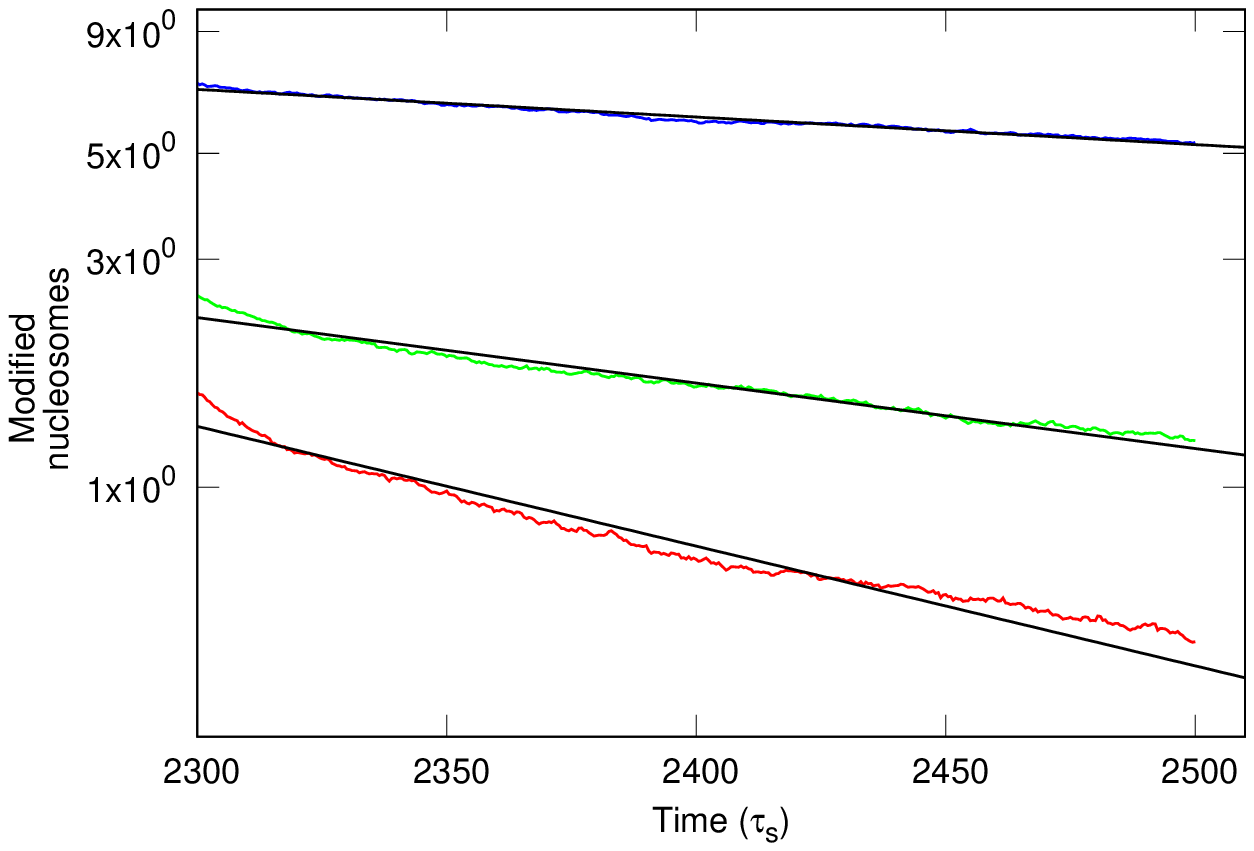} 
\end{figure} 
{\ Fits with simulations after removal of modification at certain time.} Fits of functions (black) with simulated trajectories of modified nucleosomes for sliding rates, ($k_{slide}$), 0.1 (red), 0.2 (green), 0.5 (blue) at 85\% nucleosome density. All the times are measured in units of  $\tau_{s}$. 

%\newpage
\paragraph*{Fig S3}
\label{S3_Fig}
\begin{figure}[h]
\centering
\includegraphics[width=0.60\textwidth]{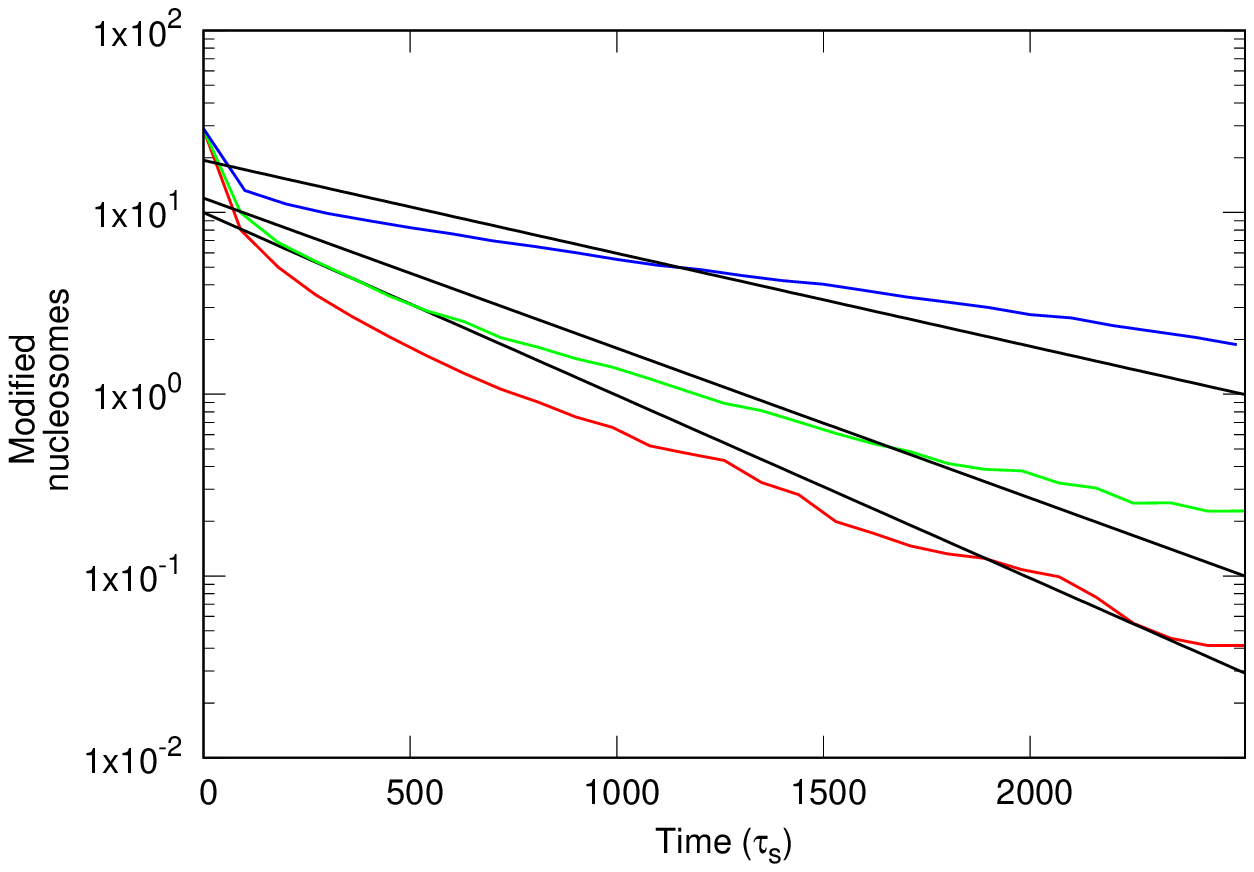} 
\end{figure} 
{\ Fits with simulations when ${\rm{N_{\rm{m}}}}$ = N.} Fits of functions (black) with simulated trajectories of modified nucleosomes when all nucleosomes 
(N) are modified at t = 0 for sliding rates, ($k_{slide}$), 0.1 (red), 0.2 (green), 0.5 (blue) at 85\% nucleosome density. All the times are measured in units of  $\tau_{s}$.

\paragraph*{Fig S4}
\label{S4_Fig}
\begin{figure}[h]
\centering
\includegraphics[width=0.60\textwidth]{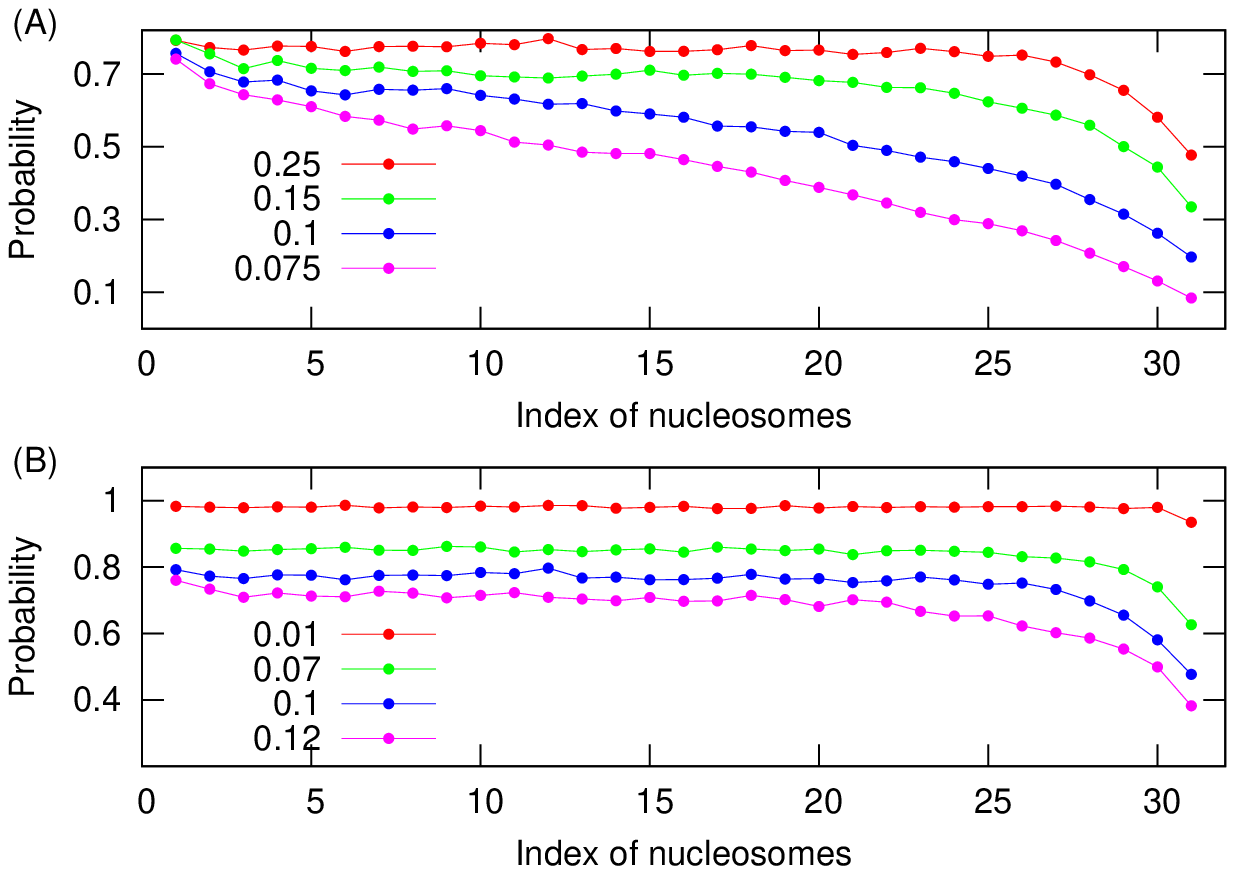} 
\end{figure} 
{\ Probabilities with sliding and de-modification rates.} The probability of modified nucleosomes at 90\% nucleosome density for sliding rates : (A) for $k_{slide}$ = 0.25 (red), 0.15 (green), 0.1 (blue) and 0.075 (magenta) and for de-modification rates (B) for $k_{r}$ = 0.01 (red), 0.07 (green), 0.1 (blue), 0.12 (magenta). All the rates are measured in units of  $\tau_{s}$.

\newpage
\paragraph*{Table S1}
\label{S1_Table}
{\ Decay constants at 90\% nucleosome density}
\begin{table}[!ht]
%\begin{adjustwidth}{-0.01in}{0in} % Comment out/remove adjustwidth environment if table fits in text column.
\centering
\begin{tabular}{l|l|l|l}
\hline
\bf{$k_{slide}$} & \bf{All modified} & \bf{Removal at t} & \bf{Fitted curve} \\ \hline
$0.1$ & $b = 7 \times10^{-5}$  & $b = 8 \times10^{-5}$ & $Y = Ae^{-bt}$ \\ \hline
$0.2$ & $b = 3 \times10^{-6}$  & $b = 9 \times10^{-6}$ & $Y = Ae^{-bt}$ \\ \hline 
$0.5$ & $b = 2.4 \times10^{-7}$  & $b = 2.4 \times10^{-7}$ & $Y = Ae^{-bt}$ \\ \hline 
\end{tabular}
%\end{adjustwidth}
\end{table}

\paragraph*{Table S2}
\label{S2_Table}
{\ Decay constants at 85\% nucleosome density}
\begin{table}[!ht]
%\begin{adjustwidth}{-0.01in}{0in} % Comment out/remove adjustwidth environment if table fits in text column.
\centering
%\caption{{\bf S2 Table Decay constants(b) at 85\% nucleosome density}}
\begin{tabular}{l|l|l|l}
\hline
\bf{$k_{slide}$} & \bf{All modified} & \bf{Removal at t} & \bf{Fitted curve} \\ \hline
$0.1$ & $b = 2.31 \times10^{-3}$  & $b = 5.77 \times10^{-3}$ & $Y = Ae^{-bt}$ \\ \hline
$0.2$ & $b = 1.90 \times10^{-3}$  & $b = 3.15 \times10^{-3}$ & $Y = Ae^{-bt}$ \\ \hline 
$0.5$ & $b = 1.17 \times10^{-3}$  & $b = 1.33 \times10^{-3}$ & $Y = Ae^{-bt}$ \\ \hline 
\end{tabular}
%\label{table1}
%\end{adjustwidth}
\end{table}